# Time-bin entangled photon pair generation from Si micro-ring resonator

Ryota Wakabayashi,[1,2] Mikio Fujiwara,[1*] Ken-ichiro Yoshino,[3] Yoshihiro Nambu,[4] Masahide Sasaki,[1] and Takao Aoki[2]

[1]*Quantum ICT Laboratory National Institute of Information and Communications Technology (NICT), 4-2-1 Nukui-kita, Koganei, Tokyo 184-8795, Japan*
[2]*Waseda University, 3-4-1 Okubo, Shinjyuku, Tokyo 165-8555, Japan*
[3]*Green Platform Research Laboratories, NEC Corporation, 1753 Shimonumabe, Nakahara-ku, Kawasaki 211-8666, Japan*
[4]*Smart Energy Research Laboratories, NEC Corporation, 34 Miyukigaoka, Tsukuba, Ibaraki 305-8501, Japan*
[*]*fujiwara@nict.go.jp*

**Abstract:** We demonstrate time-bin entanglement generation in telecom wavelength using a 7 μm radius Si micro-ring resonator pumped by a continuous wave laser. The resonator structure can enhance spontaneous four wave mixing, leading to a photon pair generation rate of about 90-100 Hz with a laser pump power of as low as -3.92 dBm (0.41 mW). We succeed in observing time-bin entanglement with the visibility over 92%. Moreover, wavelength-tunability of the entangled photon pair is demonstrated by changing the operation temperature.

## 1. Introduction

A quantum entangled photon pair source is a key component for quantum key distribution (QKD) [1], quantum metrology, and quantum information processing. Above all, QKD which allow two users to share random bits which has the unconditional security guaranteed by the fundamental laws of physics, has received much attention, because new security threats emerge these years, including advanced computers and direct attack by fiber tapping. Recently QKD systems have been deployed in the field environment and commercialized. Most widely used scheme there is the Bennett and Brassard scheme (BB84) [2] with decoy state [3,4], which can be implemented with weak pulses from a laser. Meanwhile, entanglement based QKD protocols [5] are developed as next generation methods. Entanglement based QKD has higher tolerance to side channel attacks and it does not need a random number generator, allowing simpler implementation of the system. Moreover, the distance can also be extended if qubit amplification scheme [6] is employed. The key component in practical implementation of entanglement based QKD is a compact entangled photon pair source operating at telecom wavelengths. Especially if it could be implemented in the semiconductor photonics platform, the next generation QKD systems would be more compact, and when photon detectors also are integrated, then the fully scalable quantum network can be realized.

Si waveguides [7,8] and Si micro-ring resonators [9-11] with a Si on insulator (SOI) structure, which enjoys the high index contrast between the core of Si and Silica cladding, can achieve high generation rate of photon pairs in small size. Spontaneous four wave mixing (SFWM) with $\chi^3$ nonlinear process in Si produces photon pairs. In SFWM, two pump photons are absorbed and a pair of energy- and momentum-conserving photons (named as signal and idler) are generated, saving $2\omega_p=\omega_s+\omega_p$ and $2k_p=k_s+k_i$, where $\omega_{p,s,i}$ and $k_{p,s,i}$ are the photon frequencies and wave vectors, and the subscripts *p*, *s*, and *i* are pump, signal, and idler, respectively. Using a Si waveguide, Hong-Ou-Mandel interference experiment [12] and QKD with BBM92 protocol [13] have been demonstrated [14]. Silverstone and his colleagues developed the quantum circuit which integrated a photon pair source and an interferometer in the same chip [15]. Several theoretical works [16,17] have predicted that Si micro-ring resonators are superior to Si waveguides because their cavity structures will enhance photon pair generation at the resonant wavelengths in smaller size. In addition, the bandwidth of the photon pairs emitted from Si micro-ring resonators are at least 3 orders of magnitude narrower than that from Si wire waveguides. It would help to address the entangled states and heralded single photon states to other materials. Although photon pair generations have been reported recently, there are a few works [18] for observing entanglement between photon pairs generated from a Si micro-ring resonator.

In this paper, we report the generation of entangled photon pairs from a Si micro-ring resonator in the time-bin [19-21] format, which has been firstly demonstrated in 1991 [19] and widely adopted for the fiber based QKD because of its robustness against polarization mode disturbance. Moreover, the wavelength-tunability in entangled photon pair generation is demonstrated by changing the operation temperature of the resonator. In section 2, we describe the characteristics of the resonator used in our experiment, and describe the experimental setup for measuring the photon pairs. The experimental setup for observing time-bin entanglement of photon pairs and the results are shown in section 3. Finally, we summarize our results and conclude the paper in section 4.

**2. Sample characterization and coincidence measurement**

Figure 1 shows the schematic of a micro-ring resonator coupled to two parallel waveguides. The Si micro-ring resonator with 7 μm radius horizontally couples to waveguides across 350 nm gaps. The cross section of the ring and straight line waveguides is a square of 400 nm width and 220 nm thickness. Pump light is injected into the resonator via the input waveguide and the photon pair is output through a drop port in the other waveguide (see Fig. 3). Every port of the two waveguides is connected to a single mode fiber through a spot size convertor. The transmission spectra of the resonator at temperatures of 10, 20, and 30°C are shown in Fig. 2. These data are measured by ADVANTEST Q8384 optical spectrum analyzer. The Q factor of this resonator is about 20000. The resonant wavelength can be tuned 2 nm by changing the operation temperature from 10 to 30°C. In the photon pair generation experiment, the wavelength of the pump laser is adjusted at the central peak of the resonance shown in Fig. 2.

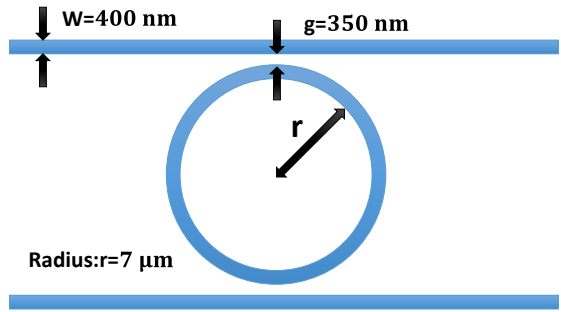

Fig. 1. Top view of the Si micro-ring resonator. The thicknesses of the ring and waveguides are 220 nm.

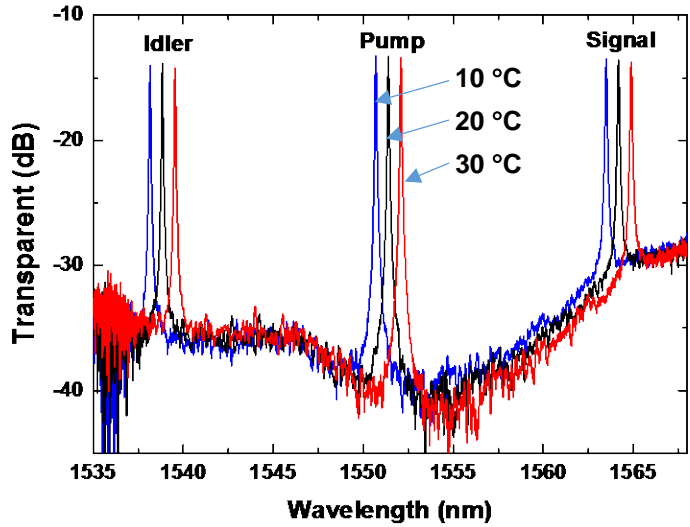

Fig. 2. Spectra of the Si micro-ring resonator at various operation temperatures (10, 20, and 30 °C).

Figure 3 shows the experimental setup for coincidence count measurement. In order to eliminate the noise in the pump light, an optical narrow bandpass filter (2 nm) is set in front of the wavelength tunable laser (Agilent 81980A). A fiber in-line module which consists of a polarizer, a half wave plate (HWP), and a quarter wave plate (QWP) is used to adjust the pump polarization to the quasi-TE mode in the micro-ring. Generated photon pairs in the resonator exit at the drop port of the waveguide. A Fiber Bragg Grating (FBG) filter is used to cut the pump light, and photon pairs are divided by a dense wavelength division multiplexing (DWDM) filter. In both ports, optical narrow bandpass filters (2 nm) are set, and photons are detected by the superconducting single photon detectors (SSPDs) [22,23]. Fiber polarization controllers are installed to compensate the polarization dependence of the detection efficiencies of the SSPDs. Electrical pulses from the SSPDs are input to the time interval analyzer (TIA [Hydra harp 400]) for analyzing coincidence counts and the time-bin entanglement.

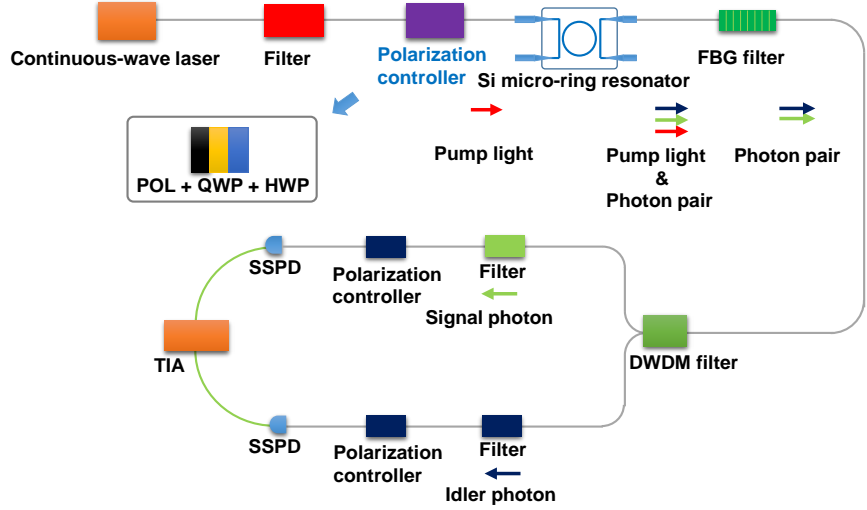

Fig. 3. Experimental setup for coincidence count measurement.

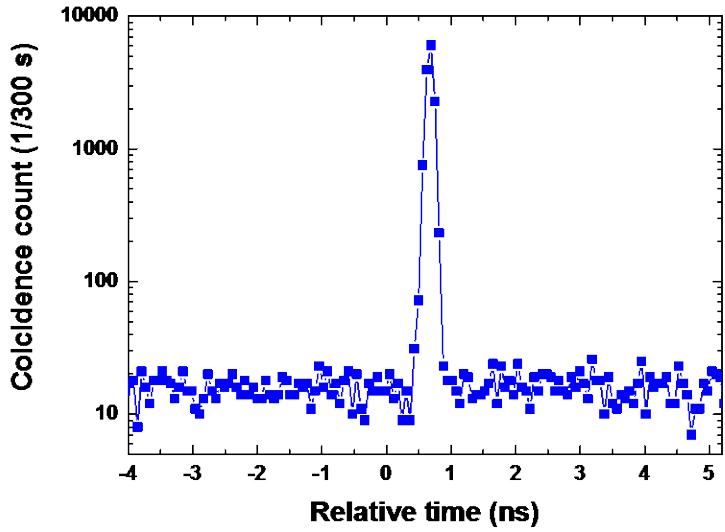

Fig. 4. Coincidence histogram. The pump power is set -3.92 dBm (0.41 mW).
The operation temperature of the Si micro-ring resonator is 25 °C.

Figure 4 shows the coincidence histogram at pump power of -3.92 dBm (0.41 mW). The coincidence count rate (CCR) is 90-100 Hz for a time window of 64 ps. The full width at half maximum (FWHM) of the histogram is estimated to be about 140 ps. This value is consistent with timing jitter of SSPDs. The coincidence to accidental ratio (CAR) is estimated to be 352±14. The CCR in this work is ten times larger than those of previous works [9-11] with the better or at least the equal of the CAR to theirs.

The transmission losses including detection losses are estimated to be 28dB for the signal and 29dB for the idler, respectively. By using these results, the intrinsic photon pair generation rate is estimated to be 21 MHz. This value is three times higher than that of the 11.3-mm-long Si straight line waveguide with the same input pump power. This result implies the ring cavity structure enables to make the highly efficient photon pair source in micrometric size.

## 3. Time-bin entanglement measurement

In the original work on the time-bin entanglement based QKD [21], the time-bin entangled state with a delay time of $\tau$ written as

$$\frac{1}{\sqrt{2}}\left\{\hat{a}^\dagger(t)\hat{b}^\dagger(t)+e^{i\theta}\hat{a}^\dagger(t+\tau)\hat{b}^\dagger(t+\tau)\right\}|0_a,0_b\rangle,$$

was generated by the double pump pulse with the delay of $\tau$. Here $\hat{a}^\dagger(t)$ and $\hat{b}^\dagger(t)$ denote the creation operators of the signal and idler photons, and $|0_a,0_b\rangle$ denotes the vacuum state of the signal and idler mode.

In our case here, the Si ring resonator is pumped by the continuous wave (CW) laser, and photon pairs mathematically written as $\hat{a}^\dagger(t)\hat{b}^\dagger(t)|0_a,0_b\rangle$ are generated in a CW beam in the drop port. Figure 5 shows the schematic of the entanglement measurement setup. Unlike with pulsed a pump laser, the time-bin entangled state is defined and made at the receivers at Alice and Bob by using an Asymmetric Mach-Zehnder Interferometer (AMZI) with the delay time of $\tau$ at each site when a CW laser used as a pump light. These AMZIs are fabricated in the planar lightwave circuit (PLC) [24] based on silica waveguide technology. The PLCs have two-inputs and four-outputs. These 2×4 AMZIs allow one to define the time-bin entanglement with the delay $\tau$ as well as to select the measurement basis set of {0,1} from the two possibilities, the X-basis or the Z-basis set. Measurement of the Z-basis corresponds to the correlation measurement of the arrival time of the photon pairs, and of the X-basis is used to measure two photon interference. Basis selection can be made by the photons themselves in the AMZIs, which is intrinsically random. Moreover, the PLC-AMZIs can operate in a polarization insensitive way by carefully adjusting the birefringence with precise temperature control. These functions allow to simplify the QKD system and enhance the resistance against Trojan horse attack [25] on the receivers. Thus our PLC-AMZIs work as time-bin entanglers and the QKD receivers, simultaneously.

The AMZI converts an operator $\hat{a}^\dagger(t)$ at the input port to $\hat{a}^\dagger(t)$, $\left\{\hat{a}^\dagger(t)\pm e^{i\theta}\hat{a}^\dagger(t+\tau)\right\}/\sqrt{2}$, and $\hat{a}^\dagger(t+\tau)$ at the four output ports Z0, X0/X1, and Z1 as shown in Fig. 5, where $\theta$ is the relative phase shifted on the long arm with respect to the short arm. Thus the Z-basis set consists of

$$|0\rangle_Z \equiv \hat{a}^\dagger(t)|0\rangle,$$
$$|1\rangle_Z \equiv \hat{a}^\dagger(t+\tau)|0\rangle,$$
(1)

while the X-basis set consists of

$$|0\rangle_X \equiv \frac{1}{\sqrt{2}}\left[\hat{a}^\dagger(t)+e^{i\theta}\hat{a}^\dagger(t-\tau)\right]|0\rangle,$$
$$|1\rangle_X \equiv \frac{1}{\sqrt{2}}\left[\hat{a}^\dagger(t)-e^{i\theta}\hat{a}^\dagger(t-\tau)\right]|0\rangle.$$
(2)

The relevant state components constituting the final entanglement between Alice and Bob are located at the slots of time, $t-\tau$, $t$ and $t+\tau$ expressed as follows;

$$|\Psi\rangle_{source} = \hat{U}_S \hat{U}_I \frac{1}{\sqrt{2}} \left[ e^{i\theta(t-\tau)} \hat{a}^\dagger(t-\tau)\hat{b}^\dagger(t-\tau) + e^{i\theta(t)} \hat{a}^\dagger(t)\hat{b}^\dagger(t) \right] |0000_s\rangle|0000_i\rangle$$

$$= \frac{e^{i\theta(t-\tau)}}{4\sqrt{2}} \left\{ \begin{array}{l} \frac{1}{\sqrt{2}}\left[\hat{a}^\dagger(t-\tau)+e^{i\theta_1}\hat{a}^\dagger(t)\right]_{X0} + \frac{1}{\sqrt{2}}\left[\hat{a}^\dagger(t-\tau)-e^{i\theta_1}\hat{a}^\dagger(t)\right]_{X1} \\ +\hat{a}^\dagger(t-\tau)_{Z0} + \hat{a}^\dagger(t)_{Z1} \end{array} \right\} |0000_s\rangle$$

$$\otimes \left\{ \begin{array}{l} \frac{1}{\sqrt{2}}\left[\hat{b}^\dagger(t-\tau)+e^{i\theta_2}\hat{b}^\dagger(t)\right]_{X'0} + \frac{1}{\sqrt{2}}\left[\hat{b}^\dagger(t-\tau)-e^{i\theta_2}\hat{b}^\dagger(t)\right]_{X'1} \\ +\hat{b}^\dagger(t-\tau)_{Z'0} + \hat{b}^\dagger(t)_{Z'1} \end{array} \right\} |0000_i\rangle$$

$$+ \frac{e^{i\theta(t)}}{4\sqrt{2}} \left\{ \begin{array}{l} \frac{1}{\sqrt{2}}\left[\hat{a}^\dagger(t)+e^{i\theta_1}\hat{a}^\dagger(t+\tau)\right]_{X0} + \frac{1}{\sqrt{2}}\left[\hat{a}^\dagger(t)-e^{i\theta_1}\hat{a}^\dagger(t+\tau)\right]_{X1} \\ +\hat{a}^\dagger(t)_{Z0} + \hat{a}^\dagger(t+\tau)_{Z1} \end{array} \right\} |0000_s\rangle$$

$$\otimes \left\{ \begin{array}{l} \frac{1}{\sqrt{2}}\left[\hat{b}^\dagger(t)+e^{i\theta_2}\hat{b}^\dagger(t+\tau)\right]_{X'0} + \frac{1}{\sqrt{2}}\left[\hat{b}^\dagger(t)-e^{i\theta_2}\hat{b}^\dagger(t+\tau)\right]_{X'1} \\ +\hat{b}^\dagger(t)_{Z'0} + \hat{b}^\dagger(t+\tau)_{Z'1} \end{array} \right\} |0000_i\rangle,$$

(3)

where $\hat{U}_s$ and $\hat{U}_i$ represent the beam splitter/combiner operations in the signal and idler modes made by the AMZIs, and $\theta(t)$ describes the relative phase of pump laser. $\theta_1$ and $\theta_2$ are the relative phase shifts between the long and short arms for the AMZIs of Alice and Bob, respectively. The subscript of creation operators corresponds to the output port as specified in Fig. 3. The subscripts $s$ and $i$ in each ket denote the signal and idler mode, respectively. The modes of the vacuum states $|0000_s\rangle$ and $|0000_i\rangle$ are understood to be (Z0, X0/X1 and Z1) and (Z'0, X'0/X'1 and Z'1), respectively, corresponding to the four output ports.

The final entanglement between Alice and Bob can be extracted by measuring the coincidence counts at the slot of time $t$, discarding the other events in the right-hand side of the second line of Eq. (3). The final entangled state detecting at the output ports X (X0, X1, X'0 and X'1) is thus

$$|\psi\rangle_{final(X\text{-basis})} = \frac{1}{2\sqrt{2}} \begin{bmatrix} \left(1+e^{i(\theta_1+\theta_2+\theta(t-\tau)-\theta(t))}\right)\hat{a}^\dagger(t)_{X0}\hat{b}^\dagger(t)_{X'0} \\ +\left(1+e^{i(\theta_1+\theta_2+\theta(t-\tau)-\theta(t))}\right)\hat{a}^\dagger(t)_{X1}\hat{b}^\dagger(t)_{X'1} \\ +\left(1-e^{i(\theta_1+\theta_2+\theta(t-\tau)-\theta(t))}\right)\hat{a}^\dagger(t)_{X0}\hat{b}^\dagger(t)_{X'1} \\ +\left(1-e^{i(\theta_1+\theta_2+\theta(t-\tau)-\theta(t))}\right)\hat{a}^\dagger(t)_{X1}\hat{b}^\dagger(t)_{X'0} \end{bmatrix} |00_s\rangle|00_i\rangle, \quad (4)$$

where the global phase factor is omitted for simplicity. $|00_s\rangle$ and $|00_i\rangle$ denote the vacuum states at the two output ports (X0 and X1) and (X'0 and X'1), respectively. The two-photon interference can be measured by changing the relative phase shifts between the long and short arms for the PLC-AMZI by controlling the operation temperature. The visibility of this interference is used to evaluate the time-bin entanglement and when it exceeds the classical limit 70.1%, one can claim that the entanglement is confirmed. By tuning the relative phase $\theta_1$ and $\theta_2$, the state can be written as

$$|\psi\rangle_{\text{final(X-basis)}} = \frac{1}{\sqrt{2}}\left(\hat{a}^{\dagger}(t)_{X0}\hat{b}^{\dagger}(t)_{X'0} + \hat{a}^{\dagger}(t)_{X1}\hat{b}^{\dagger}(t)_{X'1}\right)|00_s\rangle|00_i\rangle \quad (5)$$

in the case that $\theta_1 + \theta_2 + \theta(t-\tau) - \theta(t) = 0$, or

$$|\psi\rangle_{\text{final(X-basis)}} = \frac{1}{\sqrt{2}}\left(\hat{a}^{\dagger}(t)_{X0}\hat{b}^{\dagger}(t)_{X'1} + \hat{a}^{\dagger}(t)_{X1}\hat{b}^{\dagger}(t)_{X'0}\right)|00_s\rangle|00_i\rangle \quad (6)$$

in the case that $\theta_1 + \theta_2 + \theta(t-\tau) - \theta(t) = \pi$. As for detecting at the output ports Z (Z0, Z1, Z'0 and Z'1), the final state is

$$|\psi\rangle_{\text{final(Z-basis)}} = \frac{1}{\sqrt{2}}\left(\hat{a}^{\dagger}(t)_{Z0}\hat{b}^{\dagger}(t)_{Z'0} + e^{i(\theta(t-\tau)-\theta(t))}\hat{a}^{\dagger}(t)_{Z1}\hat{b}^{\dagger}(t)_{Z'1}\right)|00_s\rangle|00_i\rangle, \quad (7)$$

where $|00_s\rangle$ and $|00_i\rangle$ denote the vacuum states at the two output ports (Z0 and Z1) and (Z'0 and Z'1), respectively. The actual time delay is set to be $\tau$=800 ps.

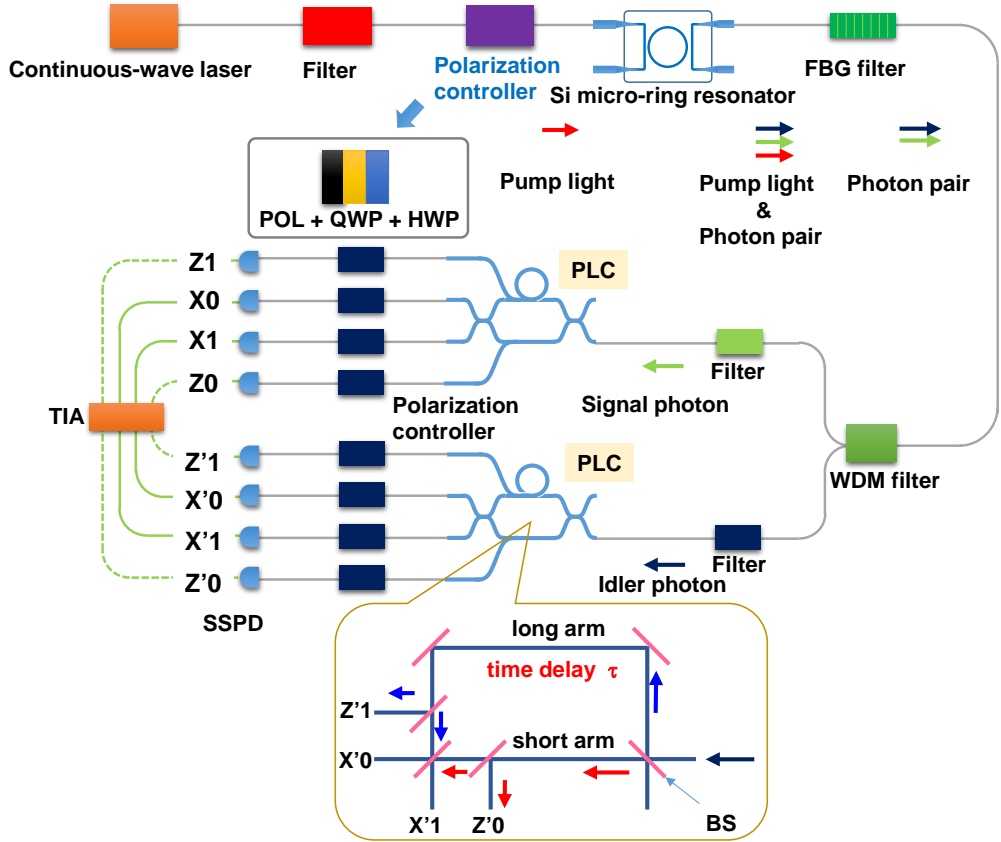

Fig. 5. Experimental setup for time-bin entanglement measurement. Inset shows the equivalent optical setup of the PLC.

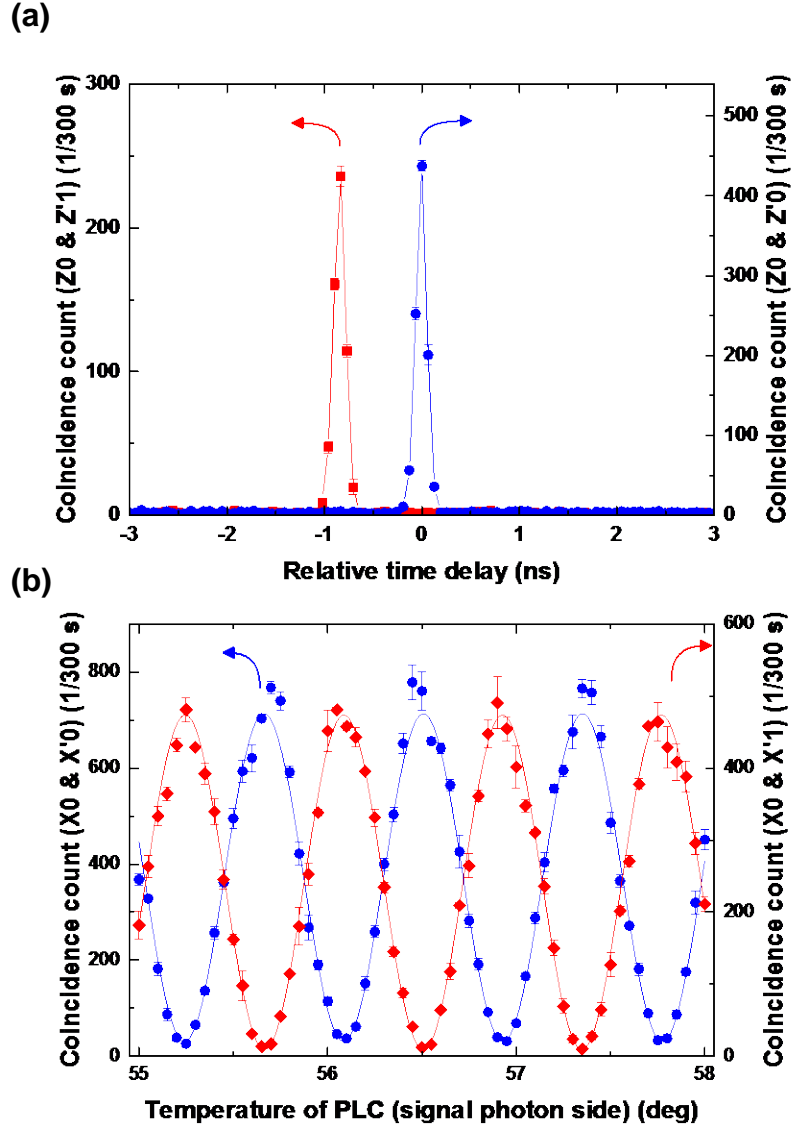

Fig. 6. (a) Coincidence histogram in Z-basis. (b) Coincidence counts in X-basis as a function of the operation temperature of the PLC (signal photon side). The time window for one data point is set 64 ps. The operation temperature of the Si micro-ring resonator is 25 °C. The input power and wavelength of the pump laser are -3.37 dBm (0.46 mW) and 1551.63 nm. The central peaks of the bandpass filters are set 1539.01 and 1564.43 nm.

For the Z-basis, the coincidence counts should be expected between Z0-Z'0 or Z1-Z'1 at the same time slot, according to Eq. (7). Coincidence counts at other ports are due to dark counts of the SSPDs, which leads to the quantum bit error in QKD. This can be estimated by evaluating the visibility defined as $\{(Z0\text{-}Z'0)-(Z0\text{-}Z'1)\}/\{(Z0\text{-}Z'0)+(Z0\text{-}Z'1)\}$. Figure 6(a) shows the coincidence counts between Z0(signal)-Z'0(idler) and Z0(signal)-Z'1(idler) (the latter should be detected at the delay time $\tau(=800$ ps) from the former). The measured visibility is 98.96±1.86%. The measured CCRs are consistent with the result shown in Fig. 4, because the transmission loss of the PLC is estimated as 7-8dB including intrinsic loss of 6dB. And this result is also consistent with the CAR obtained in Fig. 4. In this experiment, the operation temperature of the micro-ring resonator is 25 °C. The input power and wavelength are -3.37 dBm (0.46 mW) and 1551.63 nm, respectively. The central peaks of the bandpass filters for signal and idler photons are set at 1564.34 and 1539.01 nm, respectively. FWHMs of those two peaks are 140 ps, and this value is also consistent with the timing jitter of SSPDs.

The quantum interference in the X-basis is essential for time-bin entanglement. This can be measured by the coincidence counts between X0(signal)-X'0(idler), X0(signal)-X'1(idler), X1(signal)-X'0(idler) and X1(signal)-X'1(idler) by changing the relative phase shifts between the long and short arms for the PLC-AMZIs by controlling the operation temperature. The coherence time of signal and idler photons are estimated at less than 80 ps (see Fig. 2). These values are short enough to remove first order effect in the AMZI. Figure 6(b) shows the coincidence counts of X0(signal)-X'0(idler) (blue) and X0(signal)-X'1(idler) (red), respectively. We measured them for 300 seconds and three times at each point, and the error bar is by the standard deviation of the counts. The visibilities estimated by sinusoidal curve fitting are 93.22±1.15%, and 95.89±1.15% respectively. To violate the Bell's inequality, the visibility must exceed the value of the classical limit ~71% ($1/\sqrt{2}$). As a well-known value of the entanglement presentation in a quantum state and violation the Bell's inequality, parameter S [26] has been used. If S>2, the Bell's inequality is violated, and S=$2\sqrt{2}$ means the state is maximally entangled. Measured visibilities are equivalent to S=2.637±0.033 and 2.712±0.033. These results imply a clear violation of the Bell's inequality by more than 19 standard deviations.

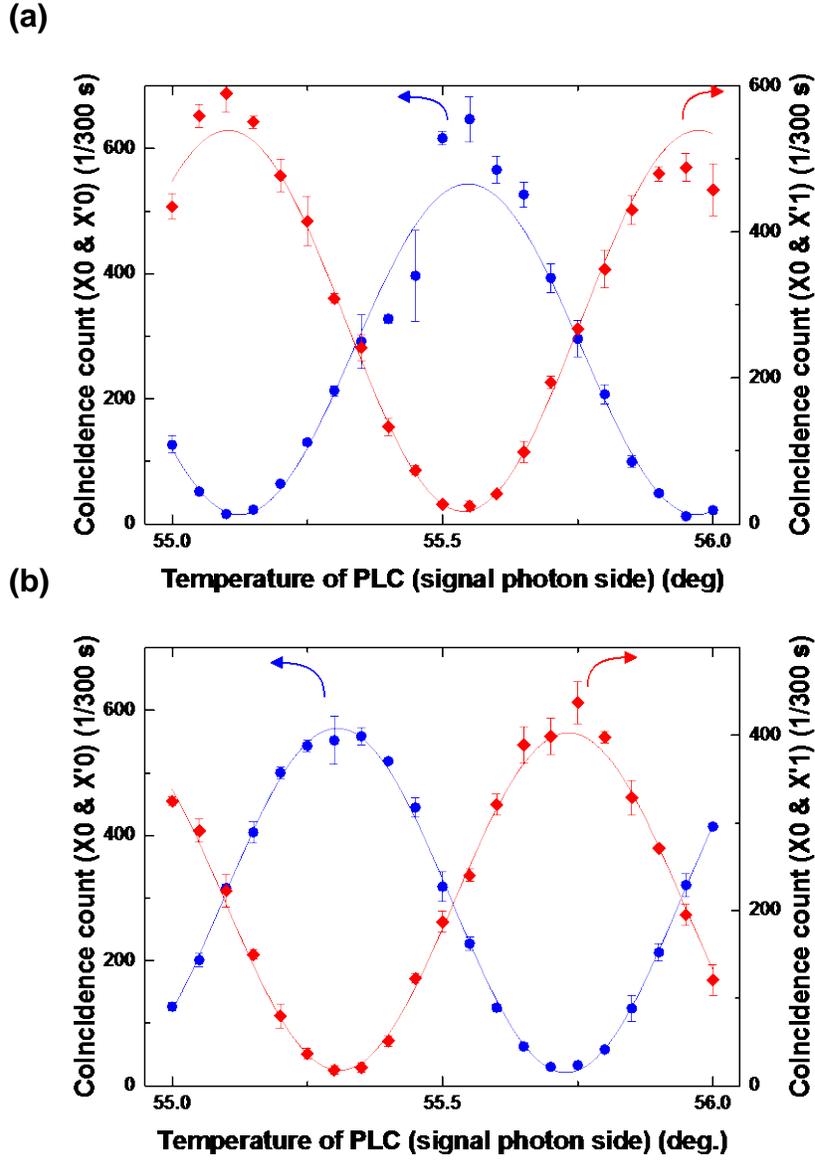

Fig. 7. Coincidence counts in X-basis as a function of the operation temperature of the PLC (signal photon side) at various operation temperature of the micro-ring. The input power of the pump laser is -3.37 dBm (0.46 mW). (a) Operation temperature is 20 °C. The wavelength of a pump laser is 1551.27 nm. Central peaks of band pass filters are 1564.05 and 1538.72 nm. (b) Operation temperature is 10 °C. The wavelength of pump laser is 1550.59 nm. Central peaks of bandpass filters are 1563.39 and 1538.04 nm.

In addition, we confirm the wavelength-tunability of the entangled photon pair by changing the temperature of the Si micro-ring resonator. The wavelength-tunability is a fundamental function to establish the entanglement swapping or other quantum experiments. Figs. 7(a) and 7(b) show the coincidence counts of the X-basis at the operation temperature of 20 and 10 °C. The wavelengths of pump laser are adjusted at 1551.27 and 1550.59 nm, respectively. The central peaks of bandpass filters for signal and idler photons are set at 1564.05, 1538.72 nm at 20 °C, and 1563.39, 1538.04 nm at 10 °C, respectively. The visibilities at both temperatures are estimated over 92% (S>2.6) which also violate the Bell's inequality. The wavelength can be tuned over a width of 2 nm in the operation temperature range of only 10-30 °C. The tunable range could be extended by expanding the operation temperature range. In addition, the difference of resonance wavelengths in 20 samples is less than 4-5 nm. To realize entanglement swapping [27] in which the indistinguishability of photons from different sources is essential feature, it is necessary to match the wavelengths of photons from different sources. The wavelength-tunability of our source would work to realize on-chip entanglement swapping or other quantum experiments by using the Si micro-ring resonators.

## 4. Conclusion

We succeed in generating time-bin entangled photon pairs from the Si micro-ring resonator with visibilities over 92%. To our knowledge, this is the first report of observing time-bin entanglement of photon pairs from the Si micro-ring resonator using PLCs which are widely used in the QKD system. This high visibility indicates the emitted photon pairs from the Si micro-ring resonator have low noise. The highly efficient photon generation is also demonstrated. In addition, we succeed in demonstrating the wavelength-tunability of the entangled photon pair by changing the operation temperature of the Si micro-ring resonator. The structure of a Si micro-ring resonator is suitable for the monolithic integration with other optical components such as optical filters and interferometers. Recently, qubit entanglement generation in an SOI platform has been reported by Silverstone et al.[28]. Their achievements would contribute to generating entangled cluster state for implementing scalable quantum information circuits. Our results show that Si micro-ring resonators would also be a useful component for making compact next generation QKD systems. Polarization entangled photon pair sources on SOI platforms were also realized [29]. A Si integrated photonic circuit allows users to generate entangled photon pairs with appropriate format according to the intended use and the characteristic of the transmission path.

**Acknowledgments**

The authors would like to thank S. Miki, T. Yamashita, and T. Terai of NICT for supplying SSPDs. And the authors appreciate M. Takeoka and E. Sasaki of NICT for their support.